\documentclass[aps,prl,reprint,showpacs,superscriptaddress,longbibliography]{revtex4-1}
\usepackage{mathtools,amssymb,graphicx,units}
\usepackage[usenames,dvipsnames]{color}
\usepackage[plainpages=false,pdfpagelabels,colorlinks=true,linkcolor=red,urlcolor=blue,citecolor=blue,pdftitle={Title},pdfauthor={},pdfdisplaydoctitle=true,pdfduplex=DuplexFlipLongEdge]{hyperref}

\graphicspath{{./Figures/}}

\begin{document}
\title{Investigating many-body mobility edges in isolated quantum systems}
\author{Xing Bo Wei}
\affiliation{Department of Physics, Zhejiang Normal University, Jinhua 321004, China}
\affiliation{Beijing Computational Science Research Center, Beijing 100193, China}
\author{Chen Cheng}
\email{chengchen@csrc.ac.cn}
\affiliation{Beijing Computational Science Research Center, Beijing 100193, China}
\author{Gao Xianlong}
\affiliation{Department of Physics, Zhejiang Normal University, Jinhua 321004, China}
\author{Rubem Mondaini}
\email{rmondaini@csrc.ac.cn}
\affiliation{Beijing Computational Science Research Center, Beijing 100193, China}

\begin{abstract}
The existence of many-body mobility edges in closed quantum systems has been the focus of intense debate after the emergence of the description of the many-body localization phenomenon. Here we propose that this issue can be settled in experiments by investigating the time evolution of local degrees of freedom, tailored for specific energies and intial states. An interacting model of spinless fermions with exponentially long-ranged tunneling amplitudes, whose non-interacting version known to display single-particle mobility edges, is used as the starting point upon which nearest-neighbor interactions are included.
We verify the manifestation of many-body mobility edges by using numerous probes, suggesting that one cannot explain their appearance as merely being a result of finite-size effects.
\end{abstract}


\maketitle

\paragraph{Introduction.---} A broad consensus on the general phenomenology of the many-body localization (MBL) is now mostly achieved~\cite{Nandkishore_15, Abanin_17, Altman_18}. It represents one of the paradigmatic examples of the novel physics that can arise from the interplay of interactions and quenched disorder in out-of-equilibrium isolated quantum systems. Specifically, it describes a new regime of the quantum matter where an emerging integrability  precludes thermalization. Since the seminal paper by Basko \textit{et al.}~\cite{Basko_06}, it has been the focus of a variety of numerical studies in different quantum models~\cite{Oganesyan_Huse_07,Znidaric_08,Pal_10,Khatami_Rigol_12,Bardarson_Pollmann_12,De_Luca_13,Iyer_13,Kjall_14,Serbym_14a,Serbym_14b,Luitz_15,Lev_15,Bera_15,Mondaini_15,Baoming_15,Torres_Herrera_15,Wang_16,Cheng_16,Serbym_16,Luitz_16,Sierant_17}, as well as in experiments of ultracold atoms~\cite{Schreiber_15,Bordia_16,Choi_16}, trapped ions~\cite{Smith_15}, solid-state spin systems~\cite{Wei_16} or in superconducting quantum processors~\cite{Martinis2017, Xu2018}. Among the more recent developments, beyond the standard scenario of static quenched disorder, studies have shown the manifestation of MBL in periodically driven quantum systems~\cite{Ponte_15, Lazarides_15, Zhang_16, Bordia_17} and even in translationally-invariant models~\cite{Grover_Fisher_14,DeRoeck_14,Schiulaz_14,Roeck_14,Schiulaz_15,van_Horssen_15,Kim_Haah_2016,Yao_Laumman_16,Mondaini_17,Smith_17}. However, a puzzling question of yet heated debate concerns the possible existence of many-body mobility edges (ME), defined as the critical energy separating localized and delocalized states. In non-interacting disordered settings, MEs are obtained in models with more than two dimensions~\cite{Abrahams_79} or in some classes of deterministic quasiperiodic systems in 1D~\cite{Biddle_10, Biddle_11, Ganeshan_15, Zhang_16, Gopalakrishnan_17, Li_17, Deng2017}.

\begin{figure}[!b] 
  \includegraphics[width=1\columnwidth]{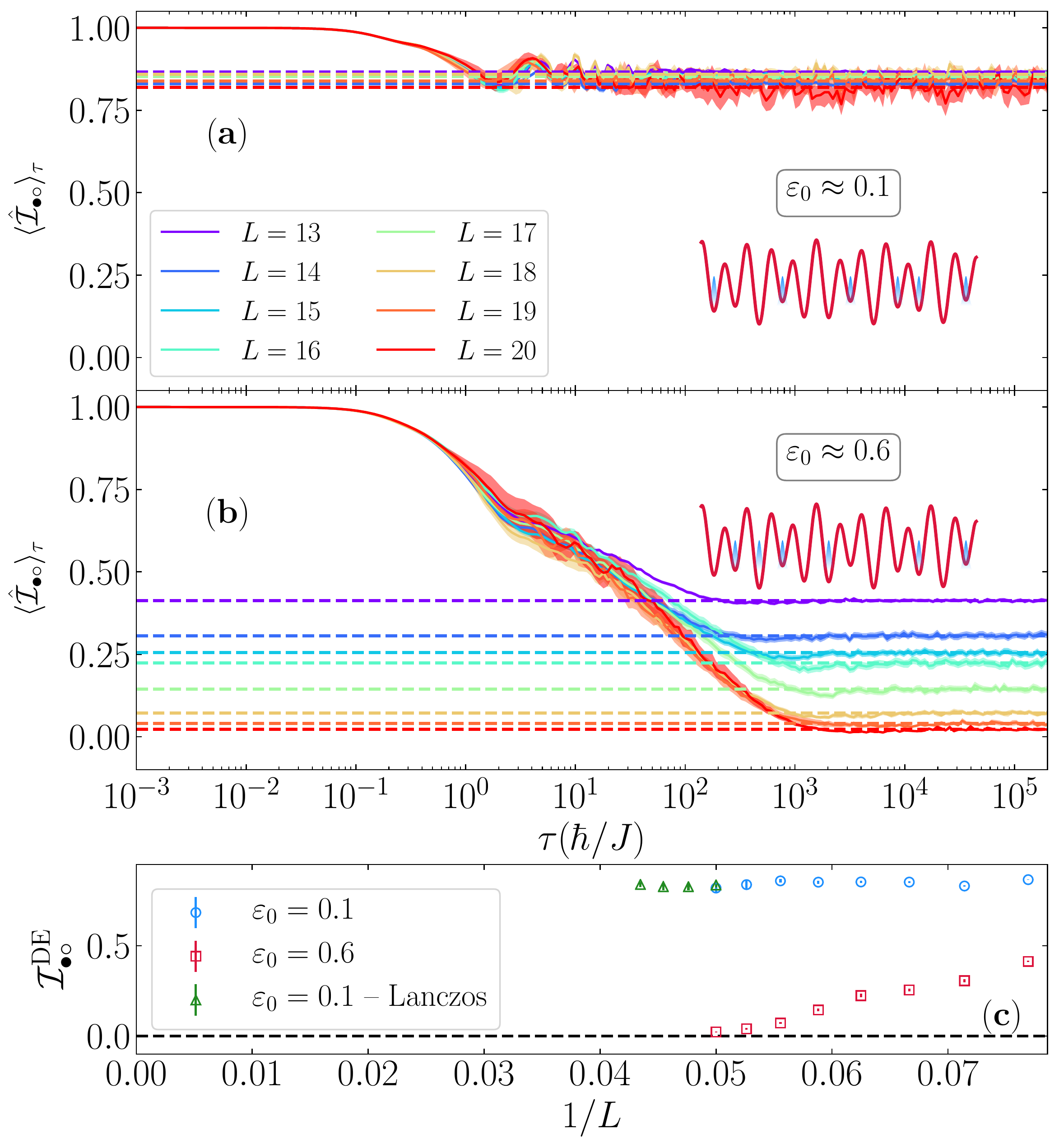}
 \vspace{-0.6cm}
 \caption{Generalized imbalance, $\hat {\cal I}_{\bullet\circ}$ (see text), and its relaxation dynamics from initial states with average total energy density $\varepsilon_0 \approx 0.1 \,[\varepsilon_0 \approx 0.6]$ in (a) [(b)], for increasing system sizes. The infinite-time average value, ${\cal I}^{\rm DE}_{\bullet\circ}$ in (c), possess a  size scaling that suggests localization in the thermodynamic limit at low energy densities, whereas for larger total energies, the system thermalizes by holding no local information of the initial preparation. The (green) triangle markers at large lattices denote the approximants obtained via the Lanczos method (see text). Shadings and error bars stem from statistical average; the cartoons in (a) and (b) depict a typical initial state whose total energy is close to the corresponding target energy density.}
 \label{fig:Fig1_PBC}
\end{figure}

One of the pioneering large scale numerical works tackling the MBL phenomenon~\cite{Luitz_15} showed that a many-body ME was manifest in a disordered one-dimensional Heisenberg model. Although system sizes attainable by exact diagonalization calculations are relatively small, this result was later corroborated by other works using techniques as, e.g., shift-and-invert MPS~\cite{Villalonga_18} and numerical linked cluster expansions which aims in directly tackling the thermodynamic limit~\cite{Devakul_15}. Despite the indication of its existence by numerics, a more recent study~\cite{DeRoeck_16} has argued that if a system displays a many-body ME, it cannot display MBL to begin with. The argument relies on the fact that highly energetic ergodic bubbles may act as a bath that leads to overall delocalization at sufficiently long times. This has sparked discussions of whether the observed phenomenon is a result of finite system sizes. Although this question is not yet solved, we believe that within our current computational capabilities, and scope of techniques available to unbiasedly obtain out-of-equilibrium properties of quantum systems, the best chance one can get in finally solving this puzzle is by providing experimentalists with a simple protocol to investigate it. This is our main goal in this Letter.

The study of (single-particle) MEs in cold atom experiments has taken renewed interest after it has been recently probed in either 3D lattices subjected to speckle disorder~\cite{Semeghini2015} or more recently in 1D optical lattices via the superposition of laser beams with different wavelengths, establishing a bichromatic quasiperiodic potential~\cite{Luschen2018}. The latter was recently employed to probe whether the resulting single-particle ME has an effect on the localization of the interacting system~\cite{Kohlert2018}. The outcome is mostly negative, in which one cannot differentiate the dynamical behavior of local observables in a system whose non-interacting limit displays or not a single-particle ME. Here, we argue that an efficient protocol to check the existence of a many-body ME, has to be modified in order to encode particularities regarding the energy of the initial state. Our conclusion is that our protocol can be used as a testbed for investigating the many-body ME in experiments, displaying its features in the long-time dynamics of local densities, as summarized in Fig.~\ref{fig:Fig1_PBC}.

Our investigation may be connected to another study~\cite{Naldesi_16} also focused on experiments: By suddenly changing the amplitudes of the onsite energies, one may pump energy into the system such that its total energy $E$ can be chosen to be above or below of the believed MEs. Here, however, we take another path which is directly connected with the recent studies in optical lattices~\cite{Luschen2018, Kohlert2018}.

\begin{figure}[!t] 
  \includegraphics[width=1\columnwidth]{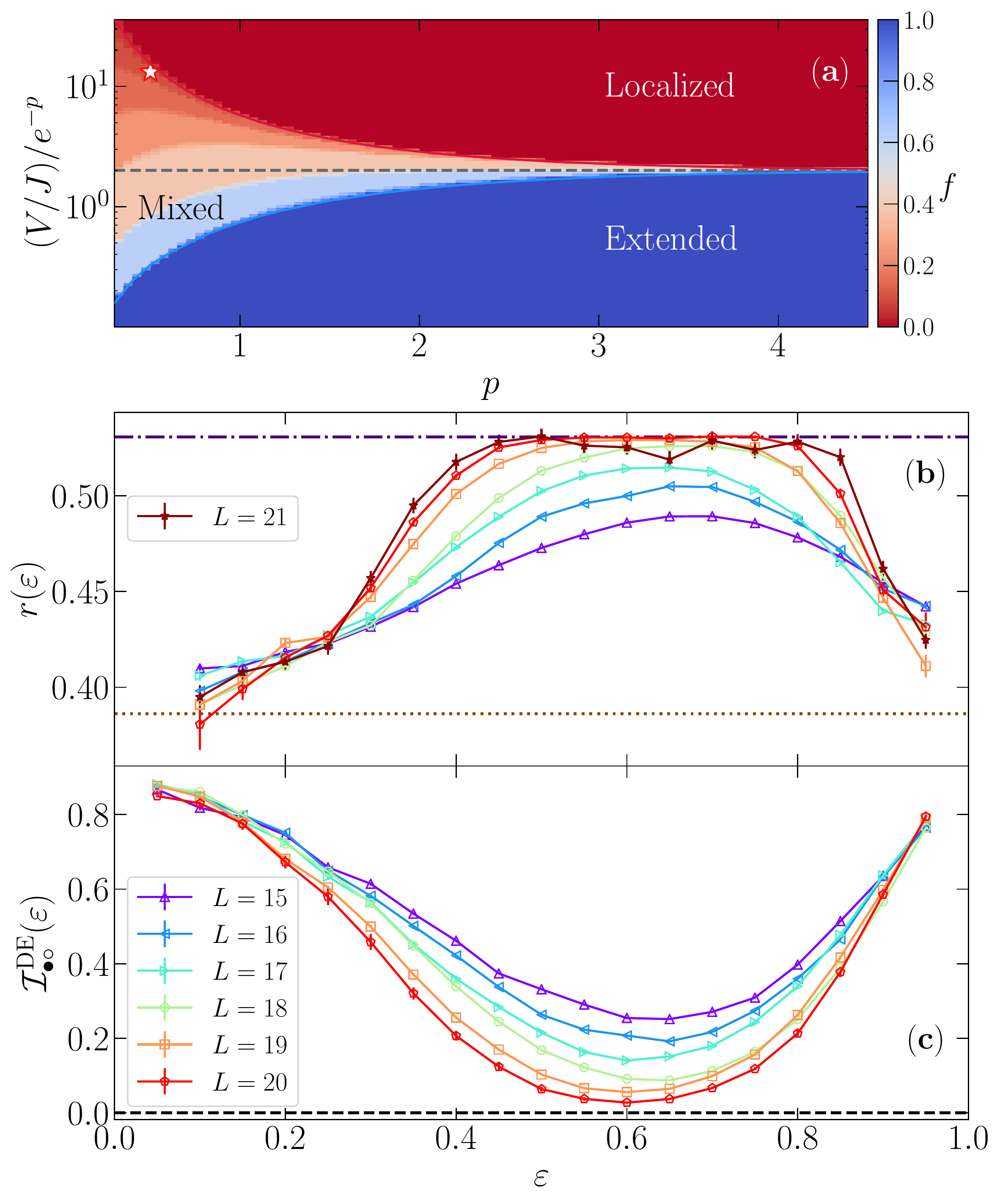}
 \vspace{-0.6cm}
 \caption{In (a), the non-interacting phase diagram of the model~\eqref{eq:hamilt} whose color denotes the fraction $f$ of extended single-particle states; the star marker depicts the point in the phase diagram we will be mostly concerned after including interactions. The dashed line marks the AA critical value, recovered in the regime $p\to\infty$ where a mixed region displaying extended and localized states is absent. In (b) and (c), the energy density resolved ratio of adjacent gaps and the infinite time average of the generalized imbalance with different $L$s. The largest lattice ($L=21$) is computed using the shift-and-invert method~\cite{petsc_user_ref, slepc_toms, Pietracaprina2018} with windows encompassing 100 eigenstates. Dashed-dotted (dotted) lines in (b) depict the limiting values of (non-)ergodic behavior.}
 \label{fig:Fig2}
\end{figure}

\paragraph{Model.---}We investigate many-body MEs being manifest in the time evolution of few-body observables in a chain with periodic boundary conditions and $L$ sites, half-filled with spinless fermions. It reads,
\begin{eqnarray}
 \label{eq:hamilt}
 \hat H = -&J&\sum_{i\neq j} e^{-p|i-j|}\left(\hat c_i^\dagger \hat c^{\phantom{}}_{j} + {\rm H.c.}\right) +U \sum_{i}\hat n_i \hat n_{i+1}\nonumber \\
 &+& V\sum_i \cos(2\pi\beta i + \varphi) \hat n_i,
\end{eqnarray}
where $\hat c_i^\dagger$ ($\hat c_i^{\phantom{}}$) creates (annihilates) a spinless fermion at site $i$ and $\hat n_i$ is a local density operator; $J$ and $V$ are the tunneling and onsite energy scales, while $U$ represents the nearest neighbor interaction magnitude. We consider the onsite energies to be emulated by a quasiperiodic potential, the Aubry-Andr\'e (AA) model~\cite{Aubry_80}, with irrational $\beta$ -- $\varphi$ allows a ``disorder'' average \footnote{We used $\beta = \frac{\sqrt{5}-1}{2}$ and typically 1500, 1000, 800, 400, 150, 134, 50 values of $\varphi$ for the system sizes $L = 15,16,17,18,19,20$, respectively, for the infinite time average in Fig.~\ref{fig:Fig1_PBC}.}--  where the hopping amplitudes are not restricted to nearest sites, but instead, have an exponentially decaying profile characterized by $p$. Its non-interacting limit ($U\to 0$) has been shown to display a single-particle ME \cite{Biddle_10}, characterized by energies $E_c = J - V\cosh{p}$, such that states with $E > E_c$ ($E<E_c$) are localized (delocalized). A full phase diagram is displayed in Fig.~\ref{fig:Fig2}(a), in terms of $p$ and the amplitude of the quasiperiodic potential $V$. Large $p$ values denote the regime where hoppings become relevant only for nearest-neighbor sites, recovering the standard AA results, where a ME is absent. We will mostly focus on the set of parameters marked by the star in this plot, i.e., $p = 0.5$ and $V /J= 8$, a regime where a fraction $f\sim7\%$ of the single-particle states are extended, considering interaction strengths $U/J = 4$.

\paragraph{Dynamically probing a many-body ME.---}\noindent Following the experiments probing MBL in optical lattices, we similarly compute the dynamics of the charge imbalance, checking whether or not it retains, at long times, information about the initial preparation. In order to understand whether ergodic behavior can coexist with localization, one needs to carefully take into account the actual total energy $E$ of the system, encoded in the initial state: $E = \langle \psi(t)|\hat H |\psi(t) \rangle = \langle \psi(0)|\hat H |\psi(0)\rangle $. A simple period-two charge density wave (CDW) state may lead to an overlook of energies where localization can be manifest. For that reason, we instead keep track of a generic local observable $\langle \hat {\cal I}_{\bullet\circ}\rangle = \rho_\bullet - \rho_\circ$, where $\rho_{\bullet,\circ} = \sum_{i\in \bullet,\circ} \langle \hat n_i\rangle/L_{\bullet,\circ}$, i.e., the difference in the densities in either the originally filled ($\bullet$) or empty ($\circ$) sites of the lattice, with $L = L_\circ + L_\bullet$. The even-odd imbalance~\cite{Schreiber_15,Bordia_16,Choi_16,Luschen2018,Kohlert2018} is a particular case of this observable which is specifically suited for the period-two CDW initial state.

We notice that the energy of the system is trivially computed via a counting problem, if the initial state closely resembles a Fock state: $E = V\sum_{i\in \bullet}\cos(2\pi\beta i + \varphi) + \sum_{i,i+1\in\bullet} U$. After reescaling the many-body spectrum as $\varepsilon = (E - E_{\rm GS})/(E_{\rm max} - E_{\rm GS})$, with $E_{\rm GS}$ ($E_{\rm max}$) being the ground-state (highest excited state), one can find, for a given realization $\varphi$ of the potential, what are the Fock states whose energy are closest to a target energy density, $\varepsilon_{0}$. Following this procedure, we report in Fig.~\ref{fig:Fig1_PBC}(a) and \ref{fig:Fig1_PBC}(b), the realization averaged dynamical imbalance for the target energy densities $\varepsilon \simeq 0.1$ and $\varepsilon \simeq 0.6$, respectively. The differences are clear: While at larger energy densities $\langle \hat {\cal I}_{\bullet\circ}\rangle_{\tau}$ steadily approaches zero for larger lattices as $\tau\to\infty$, smaller ones display a remarkable stability on the generalized charge imbalance.

One can also rule out the possibility of this being a finite-time result, since having the advantage of computing all the eigenstates $|\alpha\rangle$ of Eq.\eqref{eq:hamilt}, allows one \footnote{provided the spectrum is not plagued by degeneracies} to assess the $\tau\to\infty$ limit immediately, via the diagonal ensemble average $\langle\hat {\cal I}_{\bullet\circ}\rangle_{(\tau\to\infty)} \equiv {\cal I}^{\rm DE}_{\bullet\circ}=\sum_\alpha\langle \alpha |\hat {\cal I}_{\bullet\circ}|\alpha\rangle |\langle \psi(0)|\alpha\rangle|^2$, for an initial state $|\psi(0)\rangle$~\cite{Rigol2008, Mondaini_15, Cheng_16, Alessio_16, Mondaini_17}. These are represented by the dashed lines in Figs.~\ref{fig:Fig1_PBC}(a) and \ref{fig:Fig1_PBC}(b), showing that $\sim10^3$ hopping times ($\hbar/J$) are necessary for the system to relax and lose information of the initial density distribution at large $L$s for $\varepsilon_{0}\simeq 0.6$. Finally, we address the question of whether the apparent localization at small energy densities can be explained by the system's finiteness. In Fig.~\ref{fig:Fig1_PBC}(c), we show the size-scaling of ${\cal I}^{\rm DE}_{\bullet\circ}$: it suggests that the above mentioned stability of the generalized imbalance is maintained if one approaches the thermodynamic limit for $\varepsilon_{0} \simeq 0.1$.

\paragraph{Locating the many-body ME.---}
\begin{figure}[!t] 
  \includegraphics[width=1\columnwidth]{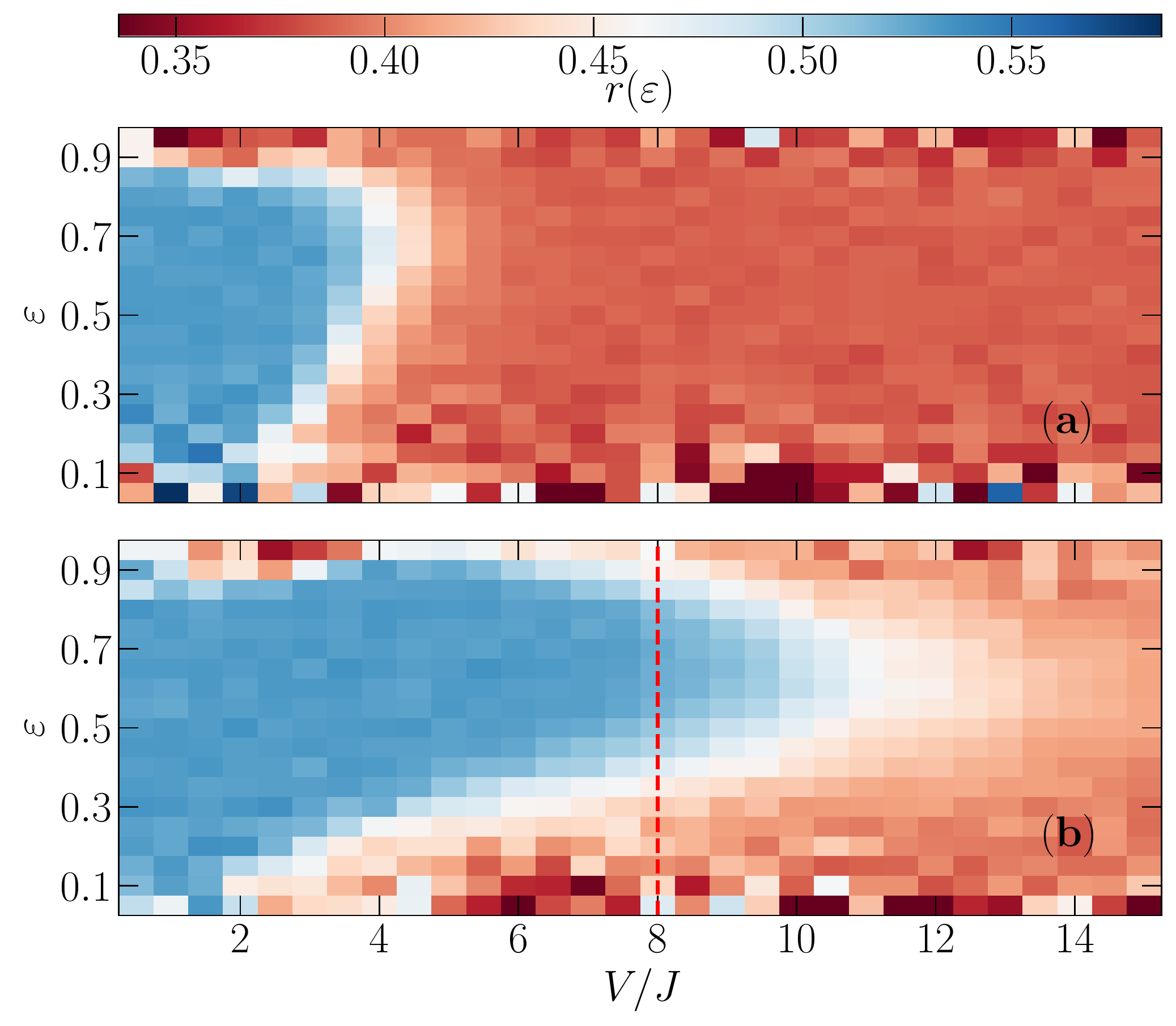}
 \vspace{-0.6cm}
 \caption{Comparison of the energy density resolved phase diagram of the ergodic and MBL regions described by the average value of the ratio of adjacent gaps in a system with $L=18$. In (a), for a model --- the same as in \eqref{eq:hamilt} but with nearest neighbor-hoppings --- without a single-particle ME in its non-interacting limit and, in (b), for our studied model that displays a single-particle ME [Eq.\eqref{eq:hamilt}]. The vertical (red) dashed line highlights the amplitude of the quasiperiodic potential we focus on.}
 \label{fig:roag}
\end{figure}

The particular values of energy density we have used are now to be justified. A generic ``belly-shape'' structure for the thermal region in the energy-disorder amplitude diagram is widely observed, either in models whose non-interacting limit displays single-particle ME or not~\cite{Kjall_14, Luitz_15, Mondragon_15, Baygan_15, Villalonga_18}. We argue here (see Fig.~\ref{fig:roag}) that the former possesses a more pronounced re-entrant behavior in the many-body localized region, signifying a more robust many-body ME -- at least in the model under consideration. Ergodic and non-ergodic regimes are characterized by the generic structure of the spectrum, quantified by the ratio of adjacent gaps~\cite{Oganesyan_Huse_07, Atas_Bogomolny_13}, $r_\alpha \equiv \min\left(\delta_{\alpha+1},\delta_{\alpha}\right)/\max\left(\delta_{\alpha+1},\delta_{\alpha}\right)$, and $\delta_\alpha = E_{\alpha}-E_{\alpha-1}$ are gaps in between consecutive energy levels in the ordered list of eigenenergies $\{E_\alpha\}$ of $\hat H$. Thermal regions manifest level-repulsion in the spectrum, whose average $\overline{r_\alpha}$ is $\approx 0.53$. In constrast, localization renders uncorrelated energy levels whose $\overline{r_\alpha}\approx 0.39$. A proper scaling analysis~\cite{Luitz_15, Modak_15, Nag_17}, shown in Fig.~\ref{fig:Fig2}(b), suggests a ME, separating ergodic from non-thermal behavior at energy densities $\varepsilon\simeq0.2$. Similarly, the infinite-time average of the generalized imbalance [Fig.~\ref{fig:Fig2}(c)] has an analogous trend, with a system-size collapse at small energy densities. Worth mentioning the possible existence of intermediate extended non-ergodic phases~\cite{Li_15, Li_16, Deng2017}, also observed in models with single-particle MEs in the non-interacting limit.

\begin{figure}[!t] 
  \includegraphics[width=1\columnwidth]{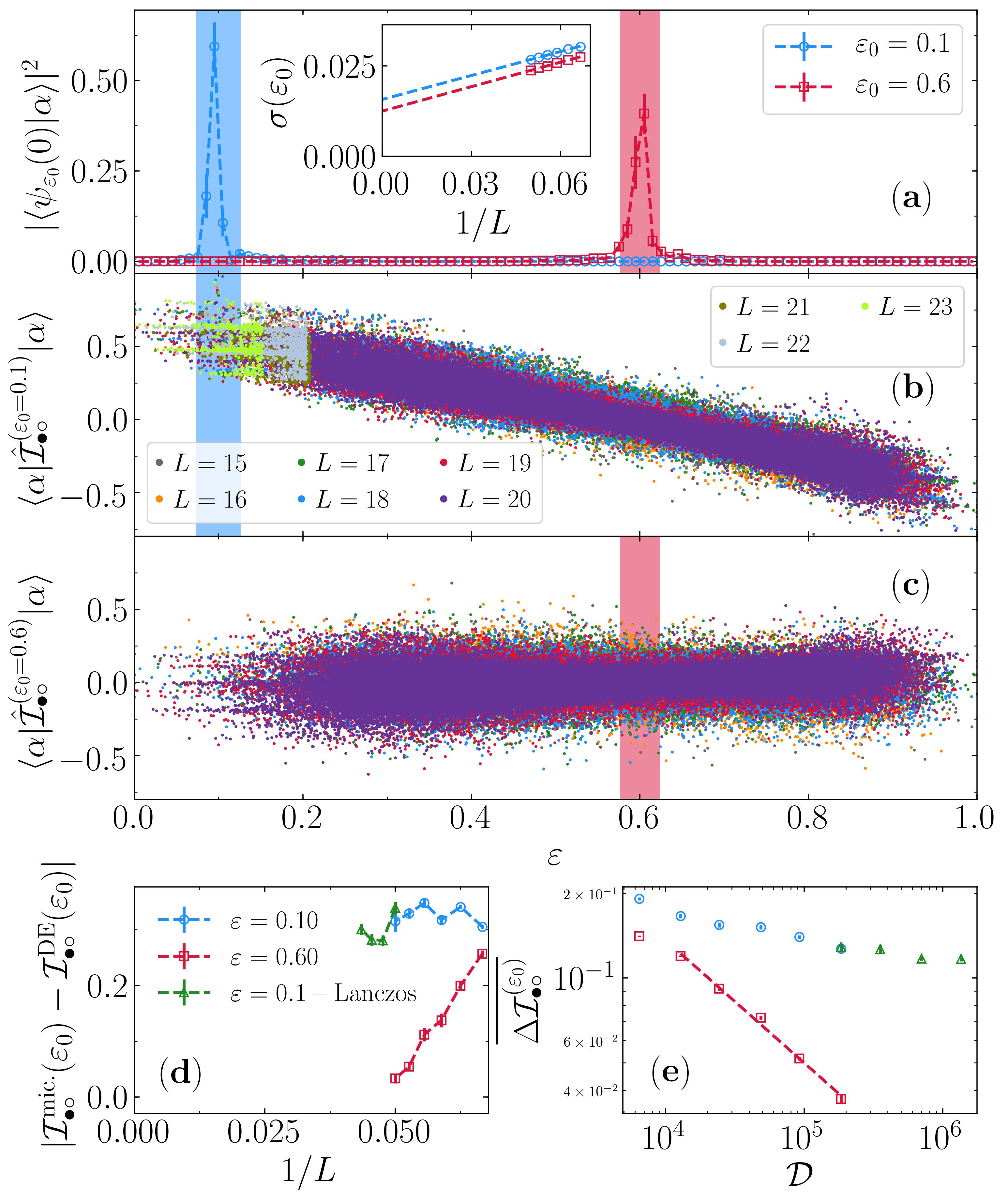}
 \vspace{-0.6cm}
 \caption{(a) Realization averaged overlap of the initial state with energy density $\varepsilon_0$ and the eigenstates of the Hamiltonian for a lattice with $L=20$; inset shows the size scaling of the width of $|\psi_{\varepsilon_0}(0)\rangle$, also represented by the shaded areas in panels [(a)-(c)]. In (b) [(c)], the eigenstate expectation value of the generalized imbalance with increasing $L$s for $\varepsilon_0 = 0.1$ [$\varepsilon_0=0.6$], including Lanczos results at large lattices. Panel (d) depicts the system size scaling of the absolute difference between the microcanonical (using $\Delta\varepsilon_0 = 0.05$) and diagonal ensembles, whereas in (e), the scaling with the Hilbert space size of the average around $\varepsilon=\varepsilon_0$ of the absolute difference between consecutive eigenstate expectation values of the generalized imbalance.}
 \label{fig:ETH_analysis}
\end{figure}

Provided the many-body ME can be thus characterized by a well defined critical energy, the dynamical protocol we describe, is only  valid in the regime where the initial state possess an energy width that does not encompasses regions with different thermal properties. This width corresponds to an estimation of how many eigenstates will effectively contribute to the dynamics of a given observable. It can be written as $\sigma(\varepsilon_0) \equiv [\langle \hat H^2\rangle - \langle \hat H \rangle^2]^{1/2}/(E_{\rm max} - E_{\rm GS})=[\sum_\alpha |\langle \alpha|\psi_{\varepsilon_0} (0)\rangle|^2(\varepsilon_0-\varepsilon_\alpha)^2]^{1/2}$~\cite{Sorg_14, Bauer_15}. The inset in Fig.~\ref{fig:ETH_analysis}(a) shows that the width of the initial (Fock) state corresponding to a given total energy is a small fraction of the bandwidth and remains so when $L\to\infty$, either in the thermal or localized region, thus guaranteeing the robustness of our protocol.

\paragraph{Thermalization predictions.---}
A final validation of the mixed character of localization that ultimately depends on the energy under investigation is given by the predictions of thermalization. A system is said to thermalize if the infinite-time average of a given observable agrees with its thermodynamic ensemble average. The eigenstate thermalization hypothesis (ETH) provides a framework for this agreement based on an ansatz for the eigenstate matrix elements of few-body observables~\cite{Rigol2008, Alessio_16}.

Figures~\ref{fig:ETH_analysis}(b) and ~\ref{fig:ETH_analysis}(c) display the diagonal matrix elements of the generalized imbalance, $\langle\alpha| {\cal \hat I}_{\bullet\circ}^{(\varepsilon_0)}|\alpha\rangle$, in the eigenstate basis when targetting the energy densities $\varepsilon_0=0.1$ and $0.6$, respectively. We start by noticing that in the former, the eigenstate expectation values are positive and remain so if increasing system sizes are considered. Thus, if one takes the definition of the diagonal ensemble, the infinite-time average of the generalized imbalance is likely to be finite (and positive), since the contributions of the negative terms in the summation is negligible, as a result of their vanishing values of $|\langle \alpha|\psi_{\varepsilon_0} (0)\rangle|^2$. This same argument, justifies using an approximant of the diagonal ensemble, where the summation is constrained to eigenstates in the vicinity of $\varepsilon_0$. By employing the Lanczos method~\cite{petsc_user_ref, slepc_toms}, one can unbiasedly get the eigenstates in this energy region and finally obtain the above approximant. Figure~\ref{fig:Fig1_PBC}(c) already includes this prediction for $\varepsilon_{0}=0.1$, where we initially show that it accurately agrees with the result of using all the eigenstates in a system with $L=20$, but later one can push to lattices with up to $L=23$. The stability of the ${\cal I}^{\rm DE}_{\bullet\circ}$ when approaching $L\to\infty$ is suggestive of localization at these energy densities. In contrast, the eigenstate expectation values of the generalized imbalance with $\varepsilon_{0} = 0.6$ [Fig.~\ref{fig:ETH_analysis}(c)] has vanishing dependence with the energy density at large lattices, with overall average of $\langle\alpha|\hat {\cal I}_{\bullet\circ}|\alpha\rangle$ close to 0. As a result, one can infer that the correspondent diagonal ensemble tends to zero, as confirmed by Fig.~\ref{fig:Fig1_PBC}(c).

We resume the connection with the ETH prediction, by noticing that the deviation between the microcanonical, ${\cal I}_{\bullet\circ}^{\rm mic.}(\varepsilon_0) \equiv \lim_{\Delta\varepsilon_0\to0} \frac{1}{{\cal N}_{\varepsilon_0}}\sum_{\alpha (|\varepsilon_\alpha-\varepsilon_0| < \Delta\varepsilon_0/2)} \langle \alpha |\hat {\cal I}_{\bullet\circ}|\alpha\rangle$~\cite{Rigol2008, Alessio_16} and the diagonal ensemble averages possess different scaling behaviors at different energy densities [Fig.~\ref{fig:ETH_analysis}(d)]. Around the middle of the spectrum, the thermalization tendencies are clear, in stark contrast with the case with $\varepsilon_0=0.1$. A final quantitative connection with the ETH is given by the eigenstate-to-eigenstate fluctuations of the eigenstate expectation values of the generalized imbalance, $\Delta {\cal I}_{\bullet\circ}^{(\varepsilon_0)}\equiv |\langle\alpha| {\cal \hat I}_{\bullet\circ}^{(\varepsilon_0)}|\alpha\rangle - \langle\alpha+1|{\cal \hat I}_{\bullet\circ}^{(\varepsilon_0)}|\alpha+1\rangle|$~\cite{beugeling_moessner_14, Kim2014, Mondaini2016}. These are known to exhibit a narrower support when approaching the thermodynamic limit if the system thermalizes, showing a scale with the dimension ${\cal D}$ of the Hilbert space as ${\cal D}^{-1/2}$~\cite{beugeling_moessner_14, Cheng_16}. Here, again resolving the behavior for different energy densities, we show in Fig.~\ref{fig:ETH_analysis}(e) that the scaling $\Delta {\cal I}_{\bullet\circ}^{(\varepsilon_0)}\propto{\cal D}^{-\gamma}$ at $\varepsilon_0 = 0.6$ possess $\gamma = 0.44(3)$. Although this is slightly off from the thermal limit, in which we attribute to the onset of Griffiths-like phases affecting even the off-diagonal matrix elements with a non-Gaussian distribution~\cite{Luitz2016} in the vicinity of the MBL regime, the system has a clear tendency to manifest thermal behavior. This is not the case at energy densities which are below the many-body ME we have inferred, where the eigenstate-to-eigenstate fluctuation has a marginal dependence on the system size.

\paragraph{Summary.---} We uncover the possibility of observing many-body MEs in experiments. Starting from a parent non-interacting Hamiltonian that possess a single-particle ME, we include interactions and analyze many probes attesting the survival of a \textit{many-body} ME. On top of different measures on the significance of thermalization below and above the critical energy density, we use a \textit{de facto} probe of MBL employed experimentally, the tracking of the time-evolved density imbalances. After a careful finite-size analysis, we argue that the existing density imbalances in the initial states survive in the long-time limit depending on the system's total energy. This energy is trivially related to the configuration of particles on the quasiperiodic potential and may be easily inferred if site addressing is available. Lastly, we point out that although our model is not exactly similar to the one in Ref.~[\onlinecite{Kohlert2018}], most prominently that the local degrees of freedom are constrained to have a maximum of one particle, the types of interactions we tackle have also been shown viable in experiments with ultracold polar molecules~\cite{Gorshkov2011}. Nevertheless, we believe that the scheme we develop here is generic and should also be visible in the emulated models with onsite interactions.

\begin{acknowledgments}
R.M. and C.C. acknowledge support from NSAF-U1530401 and from the National Natural Science Foundation of China (NSFC) Grant No. 11674021. R.M. also acknowledges support from NSFC-11650110441. G.X. and W.X. acknowlege support from NSFC under Grant Nos. 11835011 and 11774316. We acknowledge enlightening discussions with Marcos Rigol; R.M. acknowledges interactions with M. Gon\c{c}alves, E. Castro and P. Ribeiro. We thank S. Das Sarma for bringing to our attention relevant references. The computations were performed in the Tianhe-2JK at the Beijing Computational Science Research Center (CSRC).\\
\end{acknowledgments}

\bibliography{mbl_exp_hoppv1}

\end{document}